\documentclass[a4paper,11pt]{article}
\pdfoutput=1 
\usepackage{subfigure}
\usepackage{article_style} 

\title{Runaway electron velocity-space observation regions of bremsstrahlung hard X-ray spectroscopy}

\author[a,1]{E. Panontin,}
\author[a,b]{M. Nocente,}
\author[b]{A. Dal Molin,}
\author[c]{ J. Eriksson,}
\author[a,b]{G. Gorini,}
\author[a,b]{E. Perelli Cippo,}
\author[b]{D. Rigamonti,}
\author[d]{M. Salewski,}
\author[b]{M. Tardocchi}
\author[2]{and JET Contributors}

\affil[a]{Dipartimento di Fisica “G. Occhialini”, Università degli Studi di Milano-Bicocca, Milano, Italy}
\affil[b]{Istituto per la Scienza e Tecnologia dei Plasmi, Consiglio Nazionale delle Ricerche, Milano, Italy}
\affil[c]{Department of Physics and Astronomy, Uppsala University, Uppsala, Sweden}
\affil[d]{Department of Physics, Technical University of Denmark, Kgs. Lyngby, Denmark}

\begin{document}
\maketitle

\begin{center}
\noindent \corauthor{1}{e.panontin@campus.unimib.it} 
\noindent \note{2}{See the author list of "Overview of JET results for optimising ITER operation" by J. Mailloux et al. to be published in Nuclear Fusion Special issue: Overview and Summary Papers from the 28th Fusion Energy Conference (Nice, France, 10-15 May 2021)} 
\end{center}

\Abstract{The reconstruction of the distribution function of runaway electrons (RE) in magnetically confined fusion plasmas gives insights on the runaway electron beam dynamics during plasma disruptions. In view of enabling a two-dimensional, energy-pitch reconstruction of the RE velocity space, in this work we present a calculation of the weight functions for the bremsstrahlung emission by the REs. The weight functions allow bridging the bremsstrahlung spectrum with the RE velocity space, as they tell the region of the velocity space that contributes to a particular spectral measurement. The results are applied to investigate the RE velocity-space sensitivity of the hard X-ray diagnostic installed at the Joint European Torus.}

\section{Introduction}\label{sec:intro}
The study of runaway electrons (REs) in magnetically confined fusion plasmas focus on the development of suppression or at least mitigation techniques for safeguarding the structural integrity of future large devices such as ITER~\cite{gobbin2018}. The effects of such techniques on the dynamics of the RE phase-space distribution can be studied using unfolding algorithms that reconstruct the RE distribution from experimental measurements: in particular the 1D energy distribution can be retrieved from the measurements of the bremsstrahlung photons emitted in the Hard X-Ray (HXR) energy region during collisions between REs and ions in the background plasma~\cite{dalmolinPhD, panontin2021, nocente2018, shevelev2013}. 

Such an analysis is made possible on the Joint European Torus (JET) thanks to three sets of HXR diagnostics which observe the plasma on different lines of sight (LOS). A $7.62$ cm $\times$ $15.2$ cm (width × depth) cylindrical LaBr$_3$ doped Ce spectrometer (KM6T), coupled with a traditional PMT, observes the plasma tangentially~\cite{nocente2021}. Its LOS runs on the equitorial plane and spans octants from 8 to 3, intersecting the center of the poloidal plane with an angle of about $50$ deg with respect of the toroidal axis. Two spectrometers (KM6S1 and KM6S2), equipped with similar scintillators as KM6T and coupled with a traditional PMT as well, observe the plasma core vertically~\cite{nocente2013}. Finally the gamma camera upgrade (GCU) observes a poloidal section of the torus with a set of $19$ spectrometers arranged in two perpendicular arrays of $10\times9$ detectors~\cite{rigamonti2018}. Each detector is made of a $2.54$ cm × $1.69$ cm LaBr$_3$:Ce scintillator coupled to a silicon photomultiplier (SiPM)~\cite{rigamonti2018}. All the diagnostics here described are usually operated with a energy dynamic range of $[0, 30]$ MeV.

The 1D analysis of the energy distribution of REs can be extended to the 2D velocity-space thanks to the fact that the bremsstrahlung cross section depends on both the energy of the RE and the direction of emission~\cite{nocente2017, salvat2006, salvat1992}, which in turn is defined by the pitch angle and the gyro-phase of the particle. This probability is also named weight function and has been widely used for the study of fast-ions \cite{salewski2015, salewski2016} thanks to its many fold relevance. In fact, weight functions allow to study the velocity-space sensitivity of the HXR spectrometer under consideration: in particular they show to which regions the diagnostic is most sensitive to and which ones are beyond the reach. They also give hints on the contribution of the various regions of the RE 2D velocity distribution function to the measurements and they can be used to calculate synthetic measurements from RE distributions computed from first principles simulations. Finally, these weight functions are a key piece of 2D reconstruction of RE distribution; however, due to the challenges of inverting a highly undetermined system, such reconstruction will be attempted in future works. A 3D reconstruction can also be achieved by calculating orbit-based weight functions~\cite{stagner2017,jaerleblad2021,jaerleblad2022}. In this sense, the entries of the weight functions calculated in the present work represent the expected bremsstrahlung signal emitted by an RE in a certain instant of its orbit: the sum over all the RE orbit (i.e. all energy and pitch assumed by the RE during its orbit) gives a 3D orbit-based weight function. 

In the present work we will present the weight functions of the RE HXR emission, expressed as a function of energy and pitch-angle of the REs, for the HXR diagnostics installed at JET. In section~\ref{sec:WF} we introduce 2D weight functions and their computations, while in section~\ref{sec:JET} we present some examples of weight functions calculated for two Lines Of Sight (LOS) installed at JET: a tangential and a radial one. 

\section{Weight functions}\label{sec:WF}
2D velocity-space weight functions $W(E_{measured}, E_{RE}, p_{RE})$ represent the probability that an RE having energy $E_{RE}$ and pitch $p_{RE}$ will emit a photon of energy $E_{HXR}$ that will subsequently deposit an energy $E_{measured}$ in the detector. In the present work $p_{RE}=\frac{v_{RE,\parallel}}{v_{RE}}$ is defined using the same definition commonly used in fast-ions works in order to make more evident the direction of motion the the REs with respect to the magnetic field. With this definition, on JET REs mostly have negative pitch. In order to avoid confusion, it is worth underlying that in many study about REs the pitch is defined using the opposite of the RE velocity so that it can be positive.

$W(E_{measured}, E_{RE}, p_{RE})$ links the RE 2D velocity-space probability distribution function $F(E_{RE}, p_{RE})$ to the measurement of the bremsstrahlung spectrum $S(E_{measured})$ emitted by REs. In particular it defines the inverse problem:
\begin{equation}
    S(E_{m}) = \int_{LOS}n_{RE}(\underline{x})\int_0^\infty\int_{-1}^1 W(E_{m}, E_{RE}, p_{RE}, \underline{x}) F(E_{RE}, p_{RE}, \underline{x}) \,dp_{RE} \,dE_{RE} \,d\underline{x}.
\end{equation}
Here $S(E_{m})$ is the number of counts having energy in an energy bin centered in $E_{m}$ and having width $\Delta E$: $E_{measured}\in[E_{m}- \Delta E, E_{m}+\Delta E]$. $S(E_{m})$  can be conveniently normalized to the time interval over which the measurement has been performed and to the width of the energy bin $\Delta E$, in which case its units are [counts / s / MeV]. $S(E_{m})$ is the sum, performed over the detector LOS, of the signals generated by all REs having the desired energy $E_{RE}$ and pitch $p_{RE}$. In each position $\underline{x}$ inside the LOS, the number of such REs is $n_{RE}(\underline{x})F(E_{RE}, p_{RE}, \underline{x})\,dp_{RE} \,dE_{RE} \,d\underline{x}$ and the signal they generated in the detector can be calculated as $W(E_{m}, E_{RE}, p_{RE}, \underline{x}) \, n_{RE}(\underline{x}) F(E_{RE}, p_{RE}, \underline{x}) \,dp_{RE} \,dE_{RE} \,d\underline{x}$. Thus, $W(E_{measured}, E_{RE}, p_{RE}, \underline{x})$ can be seen as the spectrum generated by a mono energy and mono pitch beam, normalized to the number of REs, and its units are the same as $S(E_{m})$, in this case: [counts / s / MeV]. 

As already pointed out $W(E_{measured}, E_{RE}, p_{RE}, \underline{x})$ can be divided in two independent processes: 
\begin{equation}
    W(E_{measured}, E_{RE}, p_{RE}, \underline{x}) = 
        \int_0^\infty \, 
            W_{DRF}(E_{measured}, E_{HXR})\cdot 
            W_{B}(E_{HXR}, E_{RE}, p_{RE}, \underline{x}) \, 
        dE_{HXR}
\end{equation} 
where $W_{DRF}(E_{measured}, E_{HXR})$ is the Detector Response Function that describes the interactions between hard X-rays of energy $E_{HXR}$ and the scintillator, and it has been calculated with the MCNP code~\cite{mcnp}. On the other hand $W_{B}(E_{HXR}, E_{RE}, p_{RE}, \underline{x})$ is the emission rate of the bremsstrahlung emission and has been calculated using the GENESIS code~\cite{nocente2017}, that implements the semi-empirical cross section $\frac{d\sigma}{dE_{HXR}\,d\Omega}(E_{HXR}, E_{RE}, \theta)$ introduced in Refs.~\cite{salvat2006, salvat1992}:
\begin{equation}
W_B(E_{HXR}, E_{RE}, p_{RE}) 
        =  \frac{|\underline{v}_{RE}|}{2\pi}
            n_i\,
            d\Omega\,
            \int_0^{2\pi}
                \frac{d\sigma}{dE_{HXR}\,d\Omega}(E_{HXR}, E_{RE}, \theta(p_{RE}, \phi, \alpha))\,
            d\phi.
\end{equation}
where $|\underline{v}_{RE}|$ is the speed of the RE, $n_i$ is the ion density in the background plasma, $d\Omega$ is the solid angle subtended by the detector. The double differential cross section $\frac{d\sigma}{dE_{HXR}\,d\Omega}$ is averaged over all possible gyro-phases $\phi$ of the RE orbit in $\underline x$. In particular it is a function of the angle of emission $\theta = \arccos(\hat{LOS}\cdot\hat{v}_{RE})$, which in turn can be calculated from the particle pitch, its gyro-phase and the angle $\alpha$ between the versor $\hat{LOS}$ defining the direction of the LOS and the magnetic field.

For the sake of simplicity, in section~\ref{sec:JET} we consider a purely toroidal and uniform magnetic field, and a point-shaped RE beam positioned at the centre of the LOS under consideration. This simple situation allows us to uncover the relationship between the direction of the LOS and the magnetic field, which is the building block necessary to interpret weight functions calculated considering the full geometry of the LOS and the experimental magnetic equilibrium.

\begin{figure}[!tb]
\begin{center}
\subfigure[]{\label{img:wf:kn3g:1}
\includegraphics[scale=0.85]{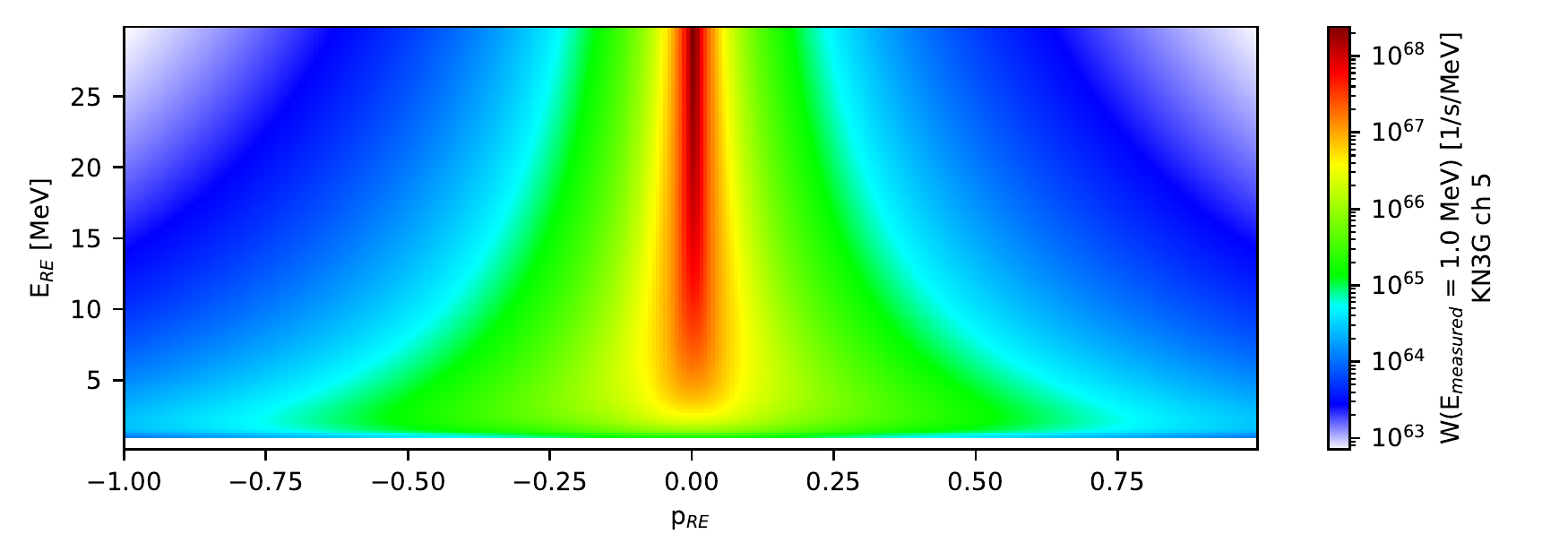}
}
\subfigure[]{\label{img:wf:kn3g:10}
\includegraphics[scale=0.85]{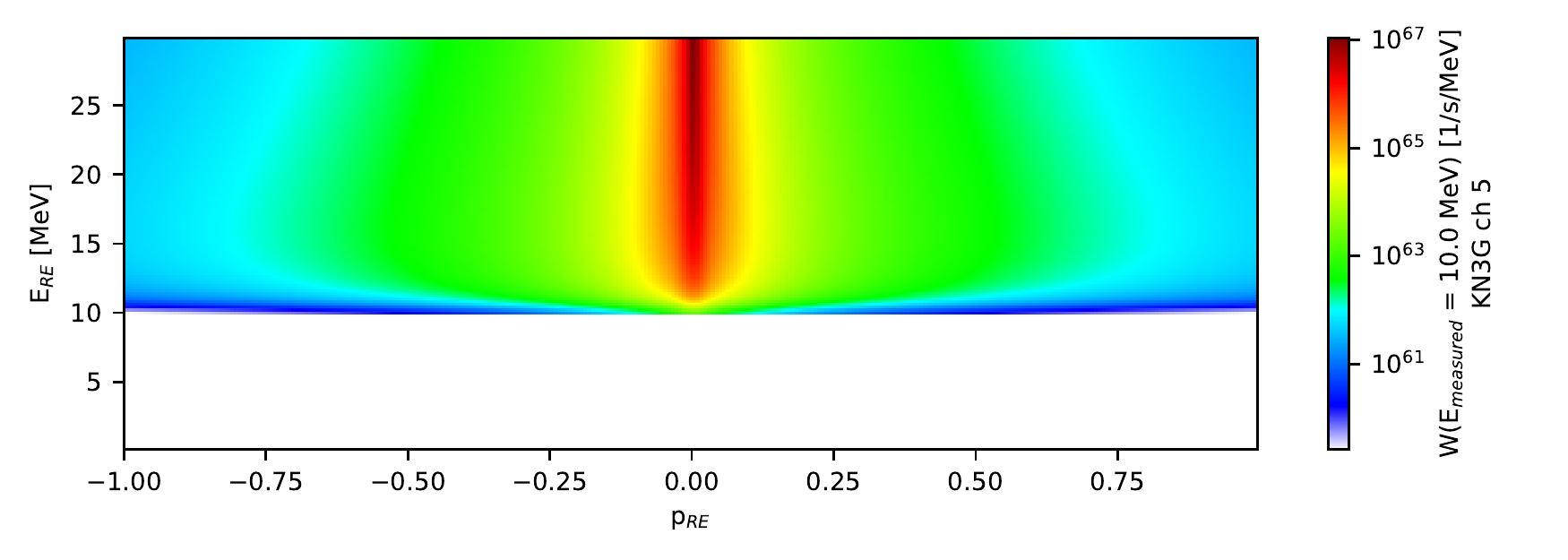}
}
\caption{\label{img:wf:kn3g} Weight functions of channel 5 of JET GCU, that has a radial LOS, considering an energy measured by the detector of $E_{measured} = 1$ MeV~\subref{img:wf:kn3g:1} and $E_{measured} = 10$ MeV~\subref{img:wf:kn3g:10}.}
\end{center}
\end{figure}

\begin{figure}[!tb]
\begin{center}
\subfigure[]{\label{img:wf:km6t:40:1:neg}
\includegraphics[scale=0.85]{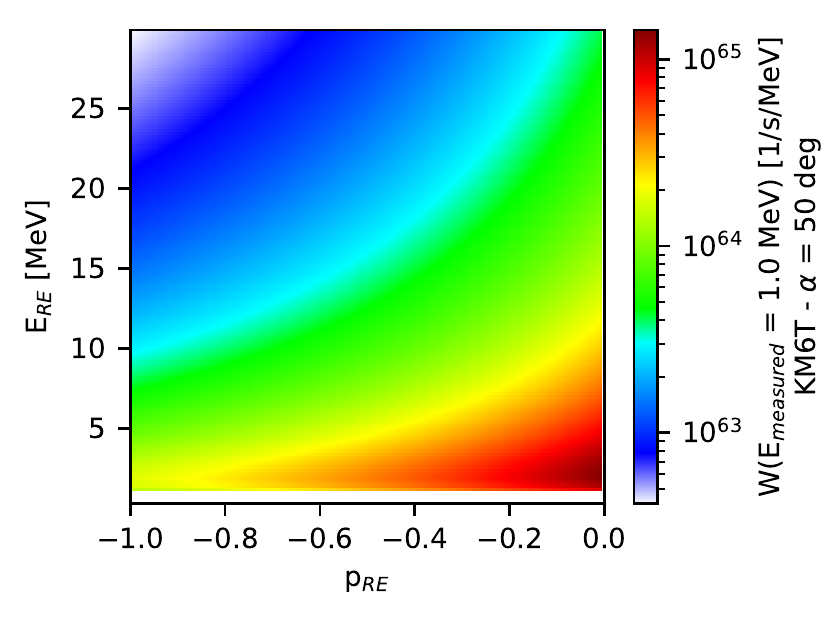}
}
\subfigure[]{\label{img:wf:km6t:40:1:pos}
\includegraphics[scale=0.85]{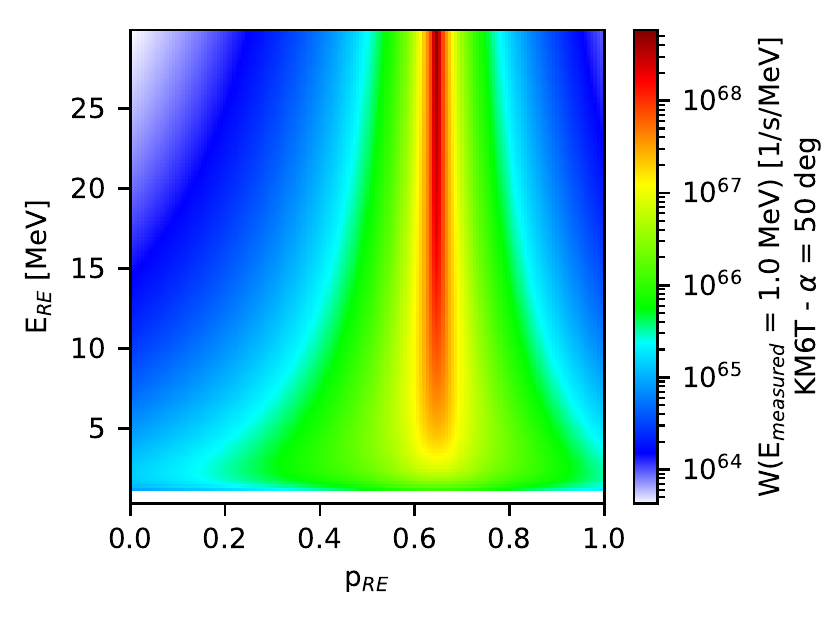}
}
\subfigure[]{\label{img:wf:km6t:40:10:neg}
\includegraphics[scale=0.85]{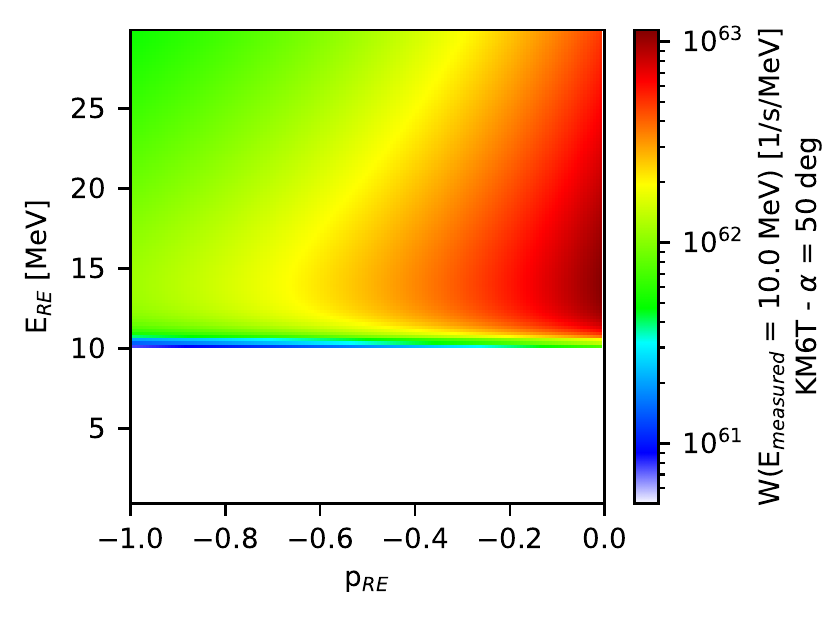}
}
\subfigure[]{\label{img:wf:km6t:40:10:pos}
\includegraphics[scale=0.85]{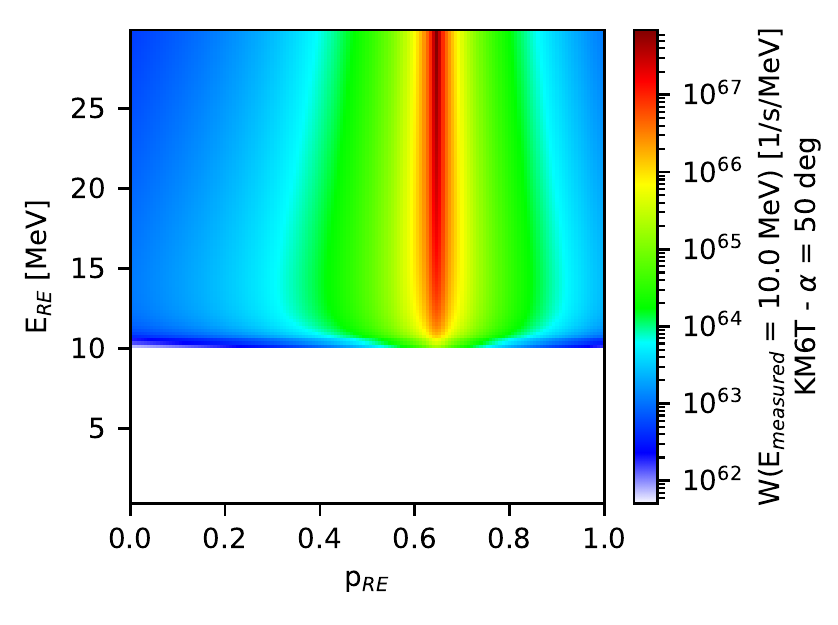}
}
\caption{\label{img:wf:km6t:40} Weight functions of the JET tangential spectrometer (KM6T) calculated for $E_{measured} = 1$ MeV~\subref{img:wf:km6t:40:1:neg}~\subref{img:wf:km6t:40:1:pos} and $E_{measured} = 10$ MeV~\subref{img:wf:km6t:40:10:neg}~\subref{img:wf:km6t:40:10:pos}. The RE beam has been placed at the center of the poloidal plane, where the LOS forms an angle $\alpha = 50$ deg with respect to the magnetic field.}
\end{center}
\end{figure}

\begin{figure}[!tb]
\begin{center}
\subfigure[]{\label{img:wf:km6t:0:1:neg}
\includegraphics[scale=0.85]{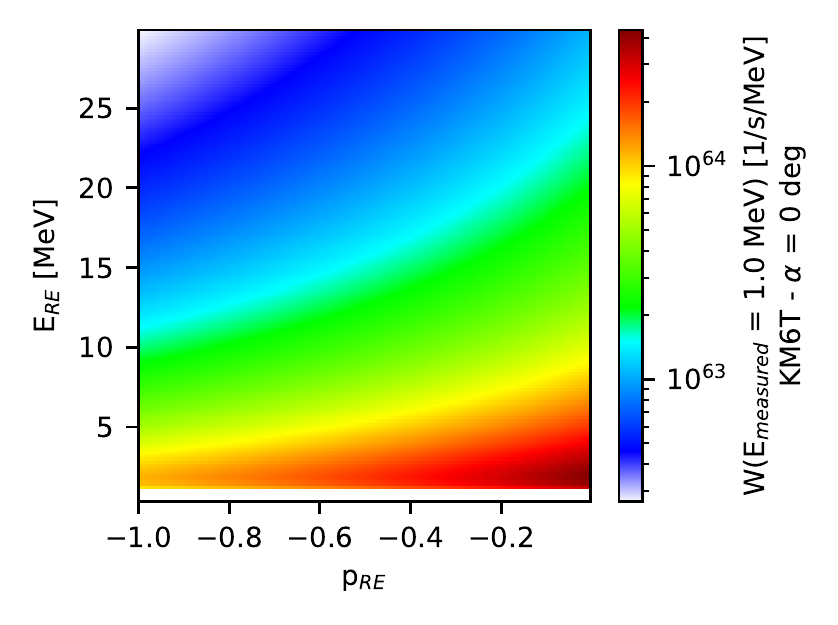}
}
\subfigure[]{\label{img:wf:km6t:0:1:pos}
\includegraphics[scale=0.85]{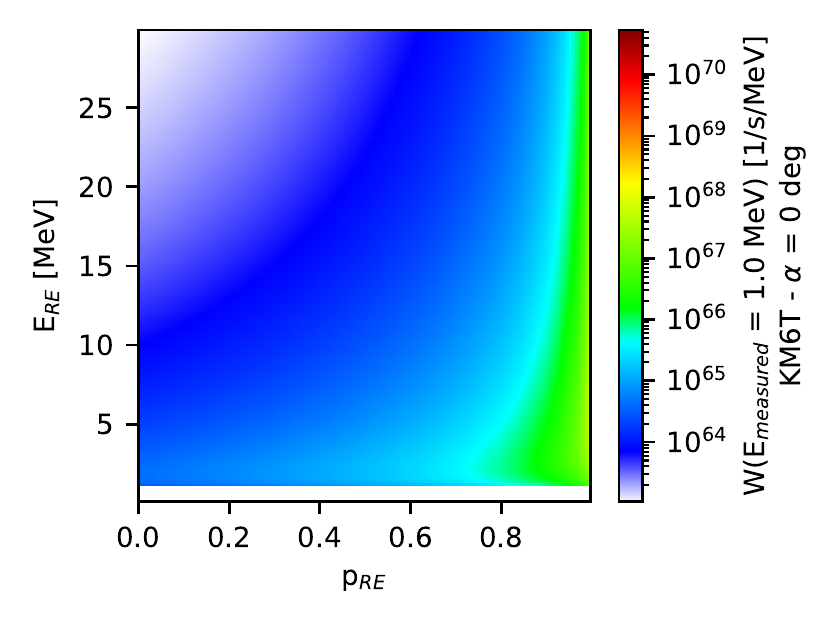}
}
\subfigure[]{\label{img:wf:km6t:0:10:neg}
\includegraphics[scale=0.85]{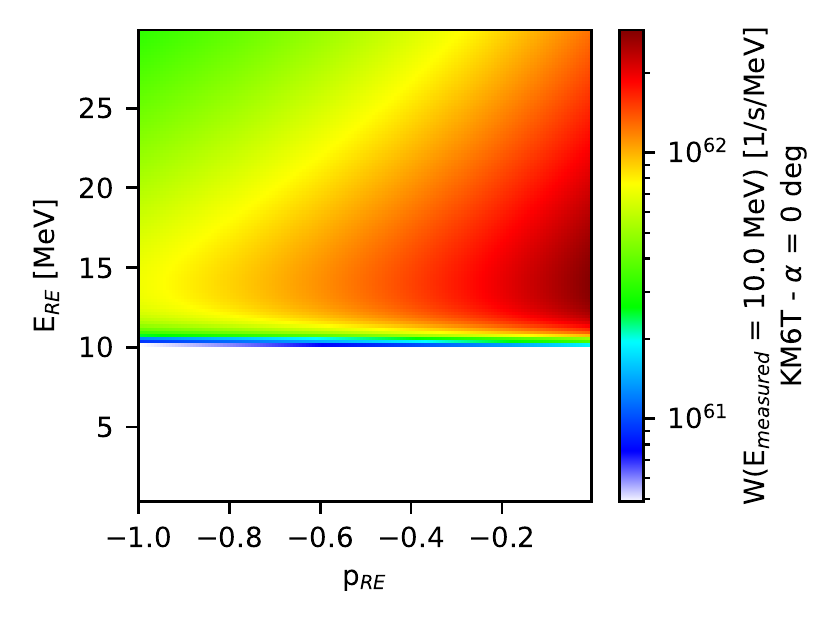}
}
\subfigure[]{\label{img:wf:km6t:0:10:pos}
\includegraphics[scale=0.85]{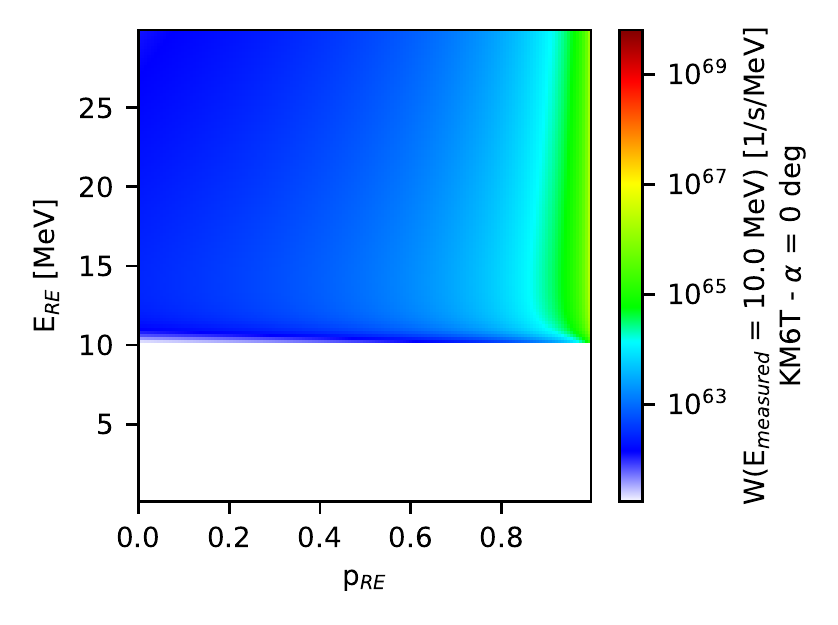}
}
\caption{\label{img:wf:km6t:0} Weight functions of the JET tangential spectrometer (KM6T) calculated for $E_{measured} = 1$ MeV~\subref{img:wf:km6t:0:1:neg}~\subref{img:wf:km6t:0:1:pos} and $E_{measured} = 10$ MeV~\subref{img:wf:km6t:0:10:neg}~\subref{img:wf:km6t:0:10:pos}. The RE beam has been placed at the high-field side, where the LOS forms an angle $\alpha = 0$ deg with respect to the magnetic field.}
\end{center}
\end{figure}

\section{Application to JET diagnostics}\label{sec:JET}
In the geometry adopted in this work (see section~\ref{sec:WF}), only the toroidal magnetic field is considered, thus all radial LOS observe the magnetic field at an angle of $90$ deg. Between one detector and another the main differences are the dimensions of the scintillator crystal and the distance of the spectrometer from the plasma, i.e. the solid angle observed by the LOS. As an example of the weight functions of radial LOS,  in figure~\ref{img:wf:kn3g} we present the 2D weight function for 1 MeV and 10 MeV signals measured by the radial LOS of channel 5 of the JET Gamma Camera Upgrade (GCU or also KN3G)~\cite{rigamonti2018}. In this particular case, particles having negative and positive pitch exhibit a symmetric behaviour and the sensitivity of the diagnostics has a peak at $p=0$. The width of this peak is inversely proportional to the energy of the particle, as in fact the more energetic the particle the more polarized the emission. The weight functions show a cutoff for $E_{RE}<E_{measured}$ as in fact the conservation of energy imposes that the energy of the emitted bremsstrahlung photon must be lower than the energy of the emitting RE. This is clearly visible in figures~\ref{img:wf:kn3g:1} and \ref{img:wf:kn3g:10}, where the weight functions for a measured energy of $1$ MeV and $10$ MeV are presented and the cutoff is at energy $E_{RE}=1$ MeV and $E_{RE}=10$ MeV respectively. As the measured energy increases, the position of the sensitivity peak remains constant and the symmetry between negative and positive pitches is also conserved. The weight functions of the other channels of the GCU and of the two vertical spectrometers are similar to the one presented in figure~\ref{img:wf:kn3g}, the only difference is in the maximum value of the weight function, which changes according to the distance of the detector from the plasma and the dimension of the crystal. 

The study of the weight functions for KM6T is made more complex by the fact that the angle $\alpha$ between its tangential LOS and the magnetic field depends on the poloidal position of the point-shaped RE beam considered for the calculations. In figure~\ref{img:wf:km6t:40}, the beam is placed at the center of the poloidal plane, where the angle $\alpha$ is about $50$ deg. Also in this case, two energies are considered: $E_{measured} = 1$ MeV and $E_{measured} = 10$ MeV. The weight functions lose the symmetry observed in figure~\ref{img:wf:kn3g} for radial LOS. In figure~\ref{img:wf:km6t:40}  the sensitivity peak is positioned at $p = \cos(50 \mbox{ deg}) = 0.64$ and the diagnostics is $3$ order of magnitudes more sensitive to particles having positive pitch (i.e. moving clockwise) than particles having negative pitch (i.e. moving counterclockwise). As before, the width of the sensitivity peak in the pitch-axis depends of the RE energy and the weight functions show a cutoff at $E_{RE}<E_{measured}$. In figure~\ref{img:wf:km6t:40}, the contributions of particles having positive pitch (figures~\subref{img:wf:km6t:40:1:pos} and~\subref{img:wf:km6t:40:10:pos}) and negative pitch (figures~\subref{img:wf:km6t:40:1:neg} and~\subref{img:wf:km6t:40:10:neg}) are reported separately with different color scales in order to include as many details as possible. The detection of the backward emission from particles having negative pitch is more probable if $E_{RE}$ is close to $E_{measured}$, this is again an effect of the anisotropy of the bremsstrahlung emission, which is more pronounced for higher $E_{RE}$. 

In many JET disruptions, the RE beam often drifts to the high-field side and the $\alpha$ between KM6T LOS and the magnetic field at the position of the RE beam decreases. The minimum value admissible is $\alpha=0$ deg, which is considered in figure~\ref{img:wf:km6t:0}. In this case the weight function is peaked at $p=1$ and the width of the peak is narrower than in figures~\ref{img:wf:kn3g} and~\ref{img:wf:km6t:40}, regardless of the measured energy or the energy of the RE. This is due to the fact that the width of the distribution of the emission angle sampled in the gyro-phase scan is proportional to $\alpha$ itself and for $\alpha=0$ deg the detector observes the photon emitted with the same $\theta$ regardless of the RE gyro-phase. Moving the position of the core of the RE beam from the high-field side to the low-field side, the angle between KM6T LOS and the toroidal magnetic field can vary from $\alpha=0$ deg to $\alpha=60$ deg. The peak in sensitivity of the weight function for KM6T, then, moves from $p=1$ to $p=0.5$ depending of the position of the RE beam and the total sensitivity of the diagnostics, calculated as $\sum_{E_{measured}, E_{RE}, p_{RE}}\underline{\underline W}$, decreases of $1$ order of magnitude.

In general, the weight function of a LOS exhibits a maximum for $p=\cos(\alpha)$, in fact for such a pitch there exists a gyro-phase for which the detector observes the photons emitted frontward, i.e. at $\theta=0$ deg. In principle one can set this peak on a desired pitch by varying the LOS inclination, thus scanning all the pitch-space.
The width of this peak depends on the angle $\alpha$ between the LOS and the magnetic field: the peak is wider for $\alpha$ close to $90$ deg and is narrower for $\alpha$ close to $0$ deg or $180$ deg. Finally it is worth noting that on JET the plasma current flows clockwise also during plasma disruptions and this means that REs have negative pitch. For example, in Refs.~\cite{nocente2018, dalmolinPhD, panontin2021} the unfolding of the RE 1D energy distribution has been conducted supposing that all REs have $p=-1$. KM6T is not very sensitive to those particles and, in order to maximise the sensitivity of a tangential LOS to the RE emission, it should have an orientation opposite to the one of KM6T. One such a LOS would also ease a 2D reconstruction of the velocity-space distribution of REs as its weight function would exhibit a peak for a negative pitch. The different position of the sensitivity peak observed in figures~\ref{img:wf:kn3g}, ~\ref{img:wf:km6t:40} and ~\ref{img:wf:km6t:0} also suggests the possibility to use multiple LOS with different orientations to discriminate between different RE orbits and calculate orbit-based weight functions, similarly to what has been recently done for fast ions in refs.~\cite{stagner2017,jaerleblad2021,jaerleblad2022}.

\section{Conclusions}
In the present work, the 2D velocity-space weight functions for channel 5 of the JET gamma camera upgrade and for JET tangential spectrometer KM6T have been presented as a function of the energy $E_{RE}$ and pitch $p_{RE}$ of the runaway electrons. In particular the weight function of the gamma camera channel is representative of all the spectrometers having radial LOS installed at JET. In the calculations presented a point-shaped runaway electron beam has been considered and the magnetic field has been set to be toroidal. The weight functions showed a peak in the diagnostics sensitivity for a pitch corresponding to the minimum angle of emission observed by the detector. It has been noted that the position and width of such a peak is determined by the angle between the LOS and the magnetic field. These matrices pave the way to a 2D reconstruction of the velocity-space distribution of runaway electrons, which will be attempted in future works.

\section*{Acknowledgements}
This work has been carried out within the framework of the EUROfusion Consortium and has received funding from the Euratom research and training programme 2014-2018 and 2019-2020 under grant agreement No 633053. The views and opinions expressed herein do not necessarily reflect those of the European Commission.

\end{document}